\documentclass[a4paper,10pt,twoside]{cpc-hepnp}

\usepackage{multicol}
\usepackage{graphicx}
\usepackage{booktabs}
\usepackage{amssymb,bm,mathrsfs,bbm,amscd}
\usepackage[tbtags]{amsmath}
\usepackage{lastpage}

\begin{document}

\fancyhead[c]{\small  Submitted to Chinese Physics
C} \fancyfoot[C]{\small \thepage}

\title{Test of the notch technique for determining the radial sensitivity of the optical model potential\thanks{Supported by National Natural Science
Foundation of China (Nos.~11375268, 11475263, U1432246 and U1432127) and the National Key Basic Research Program of China
(No.~2013CB834404) }}

\author{%
     YANG Lei
\quad LIN Cheng-Jian$^{1)}$\email{cjlin@ciae.ac.cn}%
\quad JIA Hui-Ming
\quad XU Xin-Xing
\quad MA Nan-Ru
\quad SUN Li-Jie
\quad YANG Feng \and
\quad ZHANG Huan-Qiao
\quad LIU Zu-Hua
\quad WANG Dong-Xi
}
\maketitle

\address{%
China Institute of Atomic Energy, Department of Nuclear Physics, Beijing, 102413, China
}

\begin{abstract}
Detailed investigations on the notch technique are performed on the ideal data generated by the optical model potential parameters
 extracted from the $^{16}$O+$^{208}$Pb system at the laboratory energy of 129.5 MeV, to study the
 sensitivities of this technique on the model parameters as well as the experimental data. It is found that, for the perturbation parameters, a sufficient large reduced fraction and
 an appropriate small perturbation width are necessary to determine the accurate radial sensitivity;
 while for the potential parameters, almost no dependence was observed. For the
 experimental measurements, the number of data points has little influence
 for the heavy target system, and the relative inner information of the nuclear
 potential can be derived when the measurement extended to a lower cross section.
\end{abstract}

\begin{keyword}
optical model potential; notch technique; sensitive region; elastic scattering
\end{keyword}

\begin{pacs}
24.10.Ht; 25.70.Bc
\end{pacs}

\footnotetext[0]{\hspace*{-3mm}\raisebox{0.3ex}{$\scriptstyle\copyright$}2013
Chinese Physical Society and the Institute of High Energy Physics
of the Chinese Academy of Sciences and the Institute
of Modern Physics of the Chinese Academy of Sciences and IOP Publishing Ltd}%

\begin{multicols}{2}

\section{Introduction}
\label{intro}
  The optical model potential (OMP) is the most fundamental ingredient in the study of nuclear
reaction mechanism~\cite{Brandan1997}. Nowadays, with the development of radioactive ion beams (RIBs),
the studies of the OMPs for the weakly-bound systems have attracted particular interest,
and several abnormal properties has been observed, such as the break-up threshold anomaly (BTA)~\cite{Keeley2009,Keeley2007}.

The OMP parameters can be extracted effectively by means of fitting the elastic scattering
data. However, for a given elastic scattering angular distribution, there exists numerous different
families of OMP parameters that all can give successful descriptions of the experimental data, which is the so-called Igo
ambiguity~\cite{Igo1958}. It is meaningful to discuss the OMP only within the sensitive region~\cite{Cramer1980},
where the OMP parameters can be determined accurately by the elastic scattering.
Therefore, it is quite important to know what radial regions of the nuclear potential can be well
mapped by the analysis of elastic scattering data before making any discussion on the potential.

  There are several ways to extract the radial region of the potential sensitivity~\cite{Cramer1980,Moffa1976,Lin2001}.
The frequently-used method is to find the crossing-point radius of the potential~\cite{Lin2001,Roubos2006}. However,
such a sharply-defined sensitive radius is incompatible with the principles of quantum
mechanics, and its value varies with different radial form factors adopted for the OMP~\cite{Macfarlane1981}.
Conflicting results are often brought out, e.g. the multi-crossing points~\cite{Roubos2006,Biswas2008},
especially for the energies close to the Coulomb barrier.

In Ref.~\cite{Cramer1980}, a notable technique, the notch-perturbation method was developed,
which permits an intuitionistic investigation on the sensitive region of the OMP.
Although the notch technique possesses evident advantages, only a few works~\cite{Roubos2006,Michel1983,Brown1984}
adopted this method to analyze the radial sensitivity of the OMP.
In Refs.~\cite{Cramer1980,Michel1983}, the importance of the
selection of the perturbation parameters has been suggested.
However, the dependence of this
technique on the parameters related to the perturbation, the OMP, and the experimental
measurement has not been investigated so far. In the present work,
a detailed inspection on the notch technique is performed, in order to lay a more reliable
foundation for extending the application of this technique.

\section{\label{sec:level1}The notch technique}
  The principle of the notch technique is to introduce a localized perturbation into either the real or
imaginary radial potential, and move the notch radially through the potential to investigate the
influence arising from this perturbation on the predicted cross section~\cite{Cramer1980}.

  The nuclear potential is defined as
\begin{equation}
U_{\mathrm{N}}=V(r)+iW(r)=-V_0f_{V}(r)-iW_0f_{W}(r),
\end{equation}
where the $V_0$ and $W_0$ are depths of the real and imaginary parts of the potential with
Woods-Saxon form  $f_i(r,a,R)$,
\begin{equation}
f_{i}(r,a,R)=\big [1+\exp(\frac{r-R_{i}}{a_i})\big ]^{-1}, i=V,W,
\end{equation}
where $R_i=r_{0i}(A^{1/3}_\mathrm{P}+A^{1/3}_\mathrm{T})$, $A_\mathrm{P}$ and $A_\mathrm{T}$
represent the mass numbers of the projectile and target, respectively.

  Taking the real potential $V(r)$ as an example, the perturbation of the potential
$V_\mathrm{notch}$ can be expressed as
\begin{equation}
V_{\mathrm{notch}}=dV_0f_V(R',a,R)f_{\mathrm{notch}}(r,a',R'),
\end{equation}
where, $R'$ and $a'$ represent the position and width of the notch, $d$ is the fraction by which
the potential is reduced, and $f_{\mathrm{notch}}(r, a', R')$ is the derivative Woods-Saxon
surface form factor:
\begin{equation}
f_{\mathrm{notch}}(r,a',R')=4\exp(\frac{r-R'}{a'})/[1+\exp(\frac{r-R'}{a'})]^2.
\end{equation}
Thus the perturbed real potential $V(r)_{\mathrm{pert.}}$ is:
\begin{equation}
V(r)_{\mathrm{pert.}}=V_0f_V(r,a,R)-V_{\mathrm{notch}}.
\end{equation}
  The perturbation for the imaginary potential can be derived with the same procedure.
The typical perturbed potential with $r_0 = 1.25$ fm, $a = 0.65$ fm,
$R' = 10$ fm, $a' = 0.1$ fm, and $d = 1.0$ is shown in Fig.~\ref{Fig1}.

\begin{center}
\includegraphics[width=7cm]{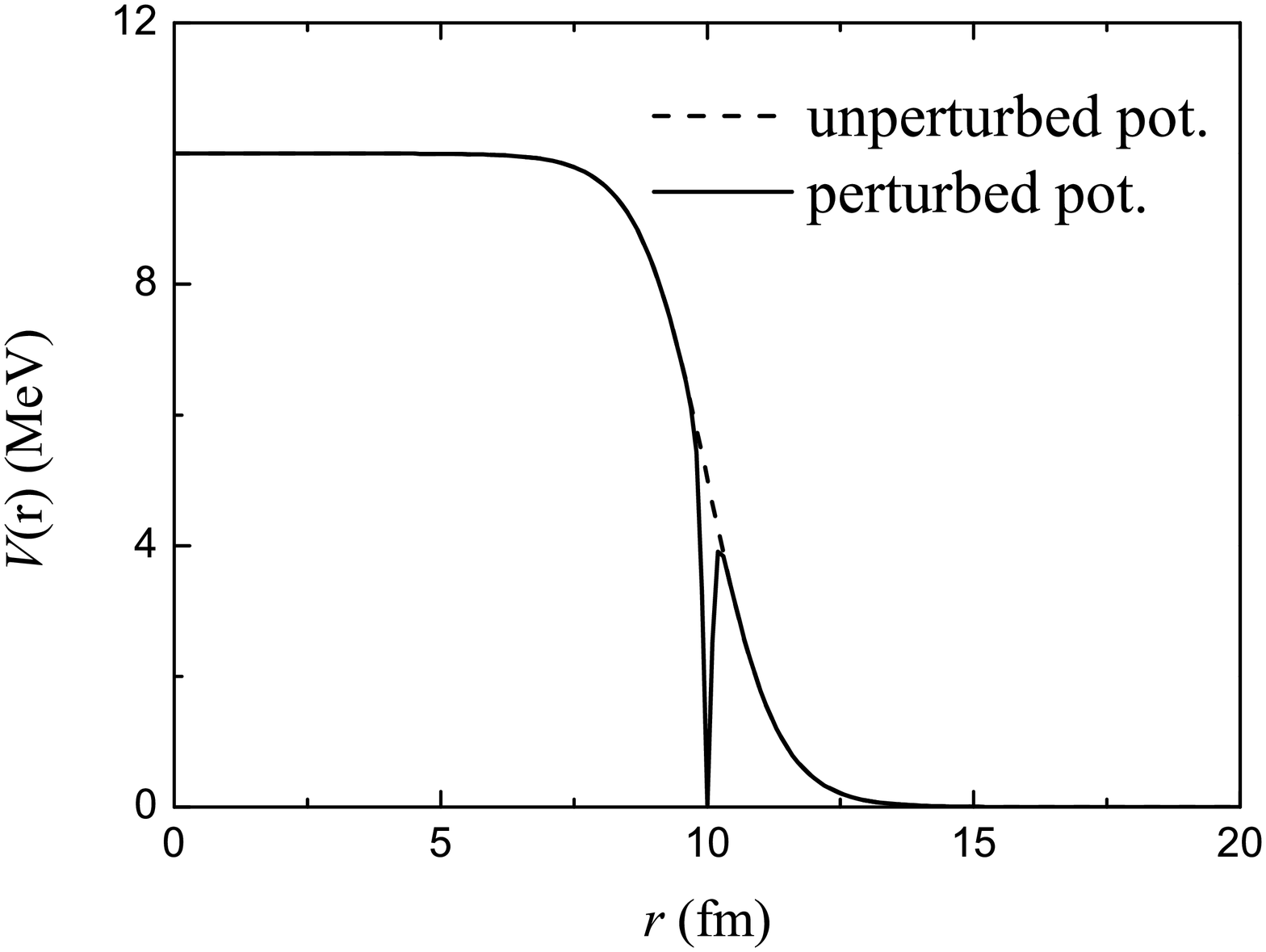}
\figcaption{\label{Fig1}   The typical unperturbed potential (dashed curve) and perturbed potential
(solid curve) with $r_0 = 1.25$ fm, $a = 0.65$ fm, $R' = 10$ fm, $a' = 0.1$ fm, and $d = 1.0$. }
\end{center}

  When the perturbation locates at the sensitive region, where the predicted cross section
depends strongly on the details of the potential, the calculated elastic scattering angular
distribution will change greatly. This means, when compared with the experimental data,
there will be a dramatic variation in the $\chi^2$ value. On the contrary, at the position where
the evaluated cross section is not sensitive to the potential, the perturbation has little
influence on the calculated angular distribution. By means of the notch technique, the sensitive region
of the nuclear potential can be presented visually and explicitly.

\section{\label{sec:level1}Sensitivity test of the notch technique}
  There may be several factors, from either the model parameters or the quality of
the experimental data, will affect the OMP sensitivity derived from the notch technique.
The influences from some possible factors will be investigated in this section, to
provide a guidance on the application of the notch technique and the procedure of experiment.
The code FRESCO~\cite{Thompson1988} was used to perform the optical model calculations.

\subsection{\label{sec:level2}data generation}

  The elastic scattering data set of $^{16}$O+$^{208}$Pb at $E_{\mathrm{lab}}(^{16}\mathrm{O}) =129.5$ MeV
~\cite{Ball1975}, as shown in Fig.~\ref{Fig2}, was chosen to perform the sensitivity test.
That is because this data is quite precise and measured in an extensive angle region
but with small angle interval, and the ratio $d\sigma_{\mathrm{el}}/d\sigma_{\mathrm{Ru}}$
was measured down to $10^{-4}$ level.
Meanwhile there is a clear picture for the interaction of this
classic tightly-bound system, and the elastic scattering angular distribution can be described
satisfactorily by the optical model. Moreover, at the energy well above the Coulomb
barrier, the nuclear force has more significant effect on the interaction progress, which
is in favor of the investigation on the radial sensitivity of nuclear potential.

\begin{center}
\includegraphics[width=7cm]{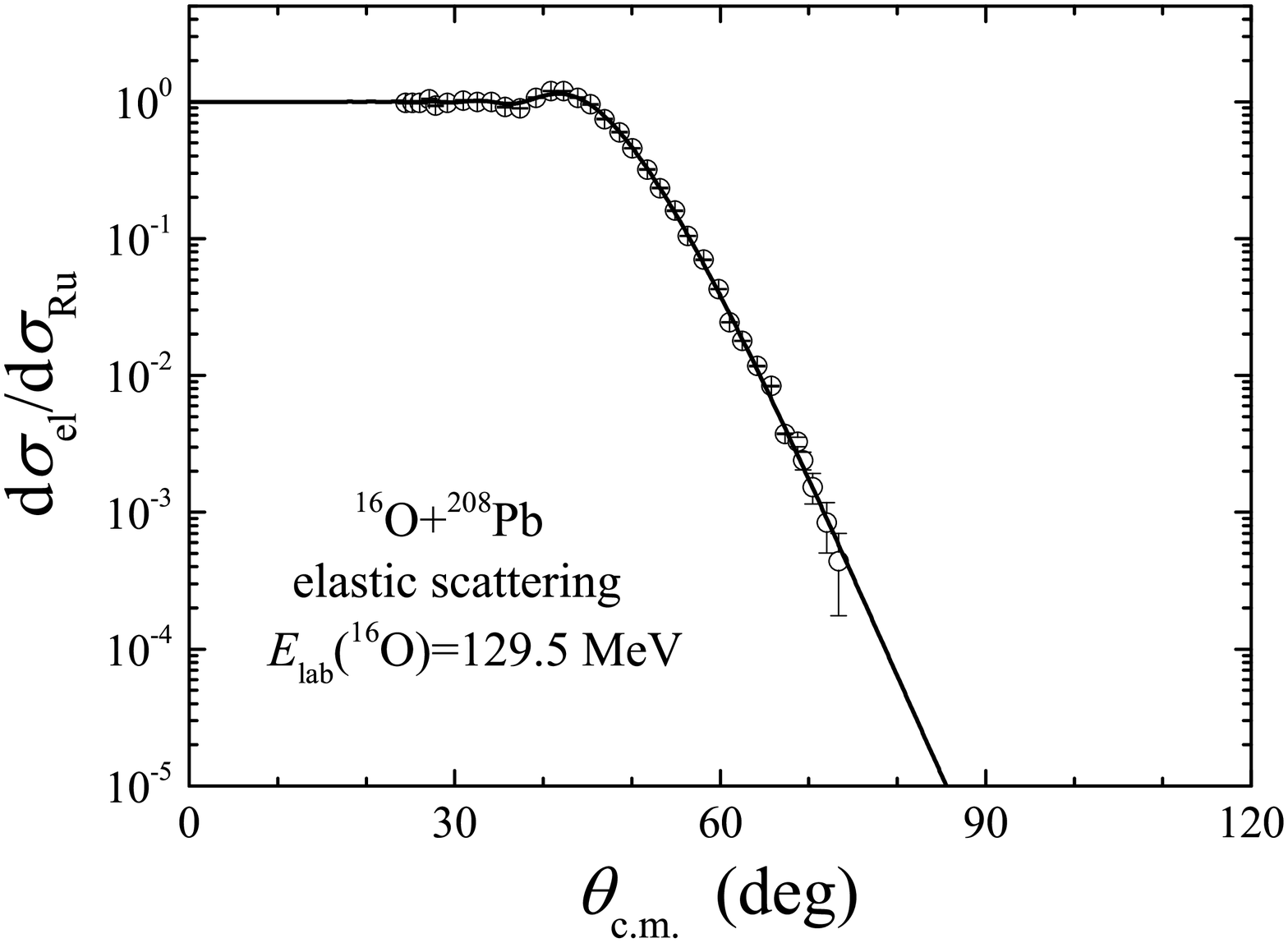}
\figcaption{\label{Fig2}   Angular distributions of $^{16}$O+$^{208}$Pb elastic scattering at
$E_{\mathrm{lab}}(^{16}\mathrm{O}) =129.5$ MeV~\cite{Ball1975}. The solid curve shows the fitting result
with $V= 31.46$ MeV, $W= 30.0$ MeV, $r_{0i} = 1.25$ fm and $a_i = 0.65$ fm, where $i= V$ and $W$.}
\end{center}

  In order to completely eliminate the uncertainties from the experimental data, such as the
statistics, angle step and range to be measured, a theoretical angular distribution
was generated by fitting the experimental data with $r_{0i} = 1.25$ fm and $a_i = 0.65$ fm,
as the solid curve shown in Fig.~\ref{Fig2}. This theoretical data can be regarded as an
ideal data set with fixed angle step of $0.1^{\circ}$ and statistic error of $1\%$.
Considering the comparability with the actual experimental
situation, the theoretical data is cut off at $\theta_{\mathrm{c.m.}} = 80^{\circ}$, where the
$d\sigma_{\mathrm{el}}/d\sigma_{\mathrm{Ru}}$ is down to $10^{-5}$ level. The following
calculations and discussions are based upon this equivalent angular distribution.

\subsection{\label{sec:level2}dependence on model parameters }

  The dependence on model parameters were investigated first, including
the perturbation parameters $d$ and $a'$, as well as the OMP parameters $r_{0i}$ and $a_i$.

\subsubsection{perturbation parameters}

  The influence of the notch depth was investigated with the value of the
reduced fraction $d$ varied in a certain step, while the notch width $a'$ fixed at 0.05 fm. The
variations of relative $\chi^2$ at different $d$ values are shown in Fig.~\ref{Fig3}. It can be seen that
the greater the perturbations are, the larger the relative $\chi^2$ values will
be brought. Distinct peaks are observed for both the real and imaginary parts,
corresponding to the radial sensitivity regions of the nuclear potential.
For the real part, a main peak lies at the position around 11.92 fm,
followed by a tiny peak in the inner region around 11.0 fm. While for
the imaginary part, two obvious peaks are observed: a major peak at around 12.20 fm
and a minor peak at around 11.20 fm. Little changes on the sensitive region are induced
by the variations of $d$, except for the the lowest $d$ value 0.2.
In that case, a broad peak was presented at about 11.5 fm in the imaginary part as shown
in the Fig.~\ref{Fig3}(b), which is incompatible with others. It indicates that a
too small reduced fraction of the perturbation may cause some spurious
sensitivity region of the potential. The $d$ value larger than 0.5 is recommended
and fixed at 1.0 in the following discussions.

\begin{center}
\includegraphics[width=7cm]{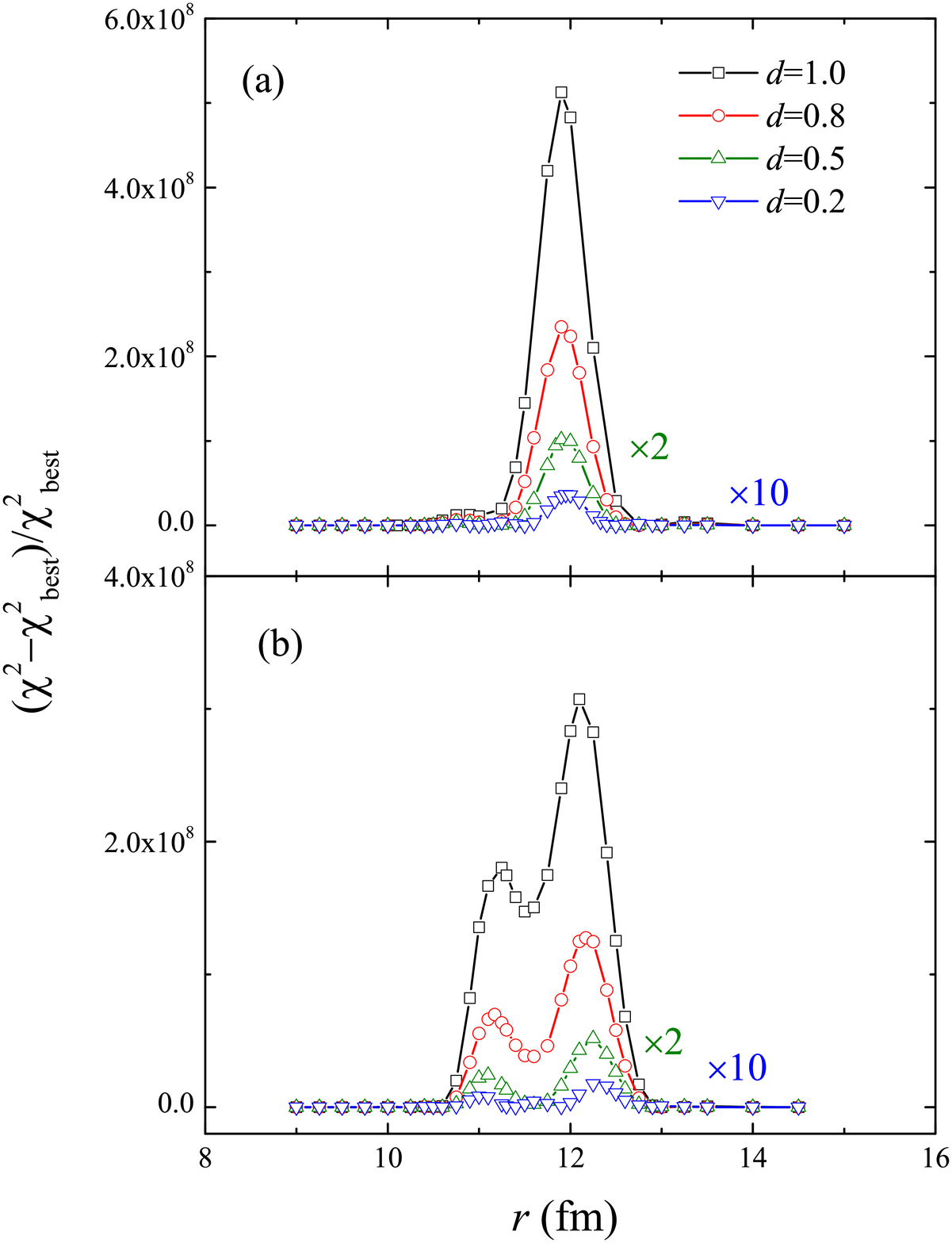}
\figcaption{\label{Fig3}   (Color online) The sensitivity functions for the real part (a) and imaginary part (b)
potentials with different $d$ values. The ``$\times10$" symbols mean the corresponding results with the
same color are multiplied by 10 for the convenience of comparison. There are same meanings for
times signs in the following figures. The curves are used to guide the eyes.}
\end{center}

  The notch width $a'$ was set at 0.2, 0.1, 0.05 and 0.01 fm, respectively, and the corresponding
sensitivity functions are shown in Fig.~\ref{Fig4}. It can be seen that a wider perturbation
introduces a stronger influence, leading to a larger relative $\chi^2$ value.
However, the original distinct double-peaked structure in the imaginary part disappears when a large
value of $a'$ was adopted, replaced by one broad peak containing the gross
information of the radial sensitivity. On the contrary, when
$a' =0.01$ fm, which is equal to the integration step size $\mathrm{d}r$, three peaks emerge
in the sensitivity function of the imaginary part potential. As mentioned in
Ref.~\cite{Cramer1980}, problems may be arising when
$a'$ becomes comparable to the $\mathrm{d}r$. Based on the above discussions, we argue that
a smaller $a'$ is benefit to extract the fine information on the sensitivity of the radial
potential, but not too close to the integration step. The $a'$ value of 0.05 fm, about 5 times
of the integration step $\mathrm{d}r$, is adopted afterward.

\begin{center}
\includegraphics[width=7cm]{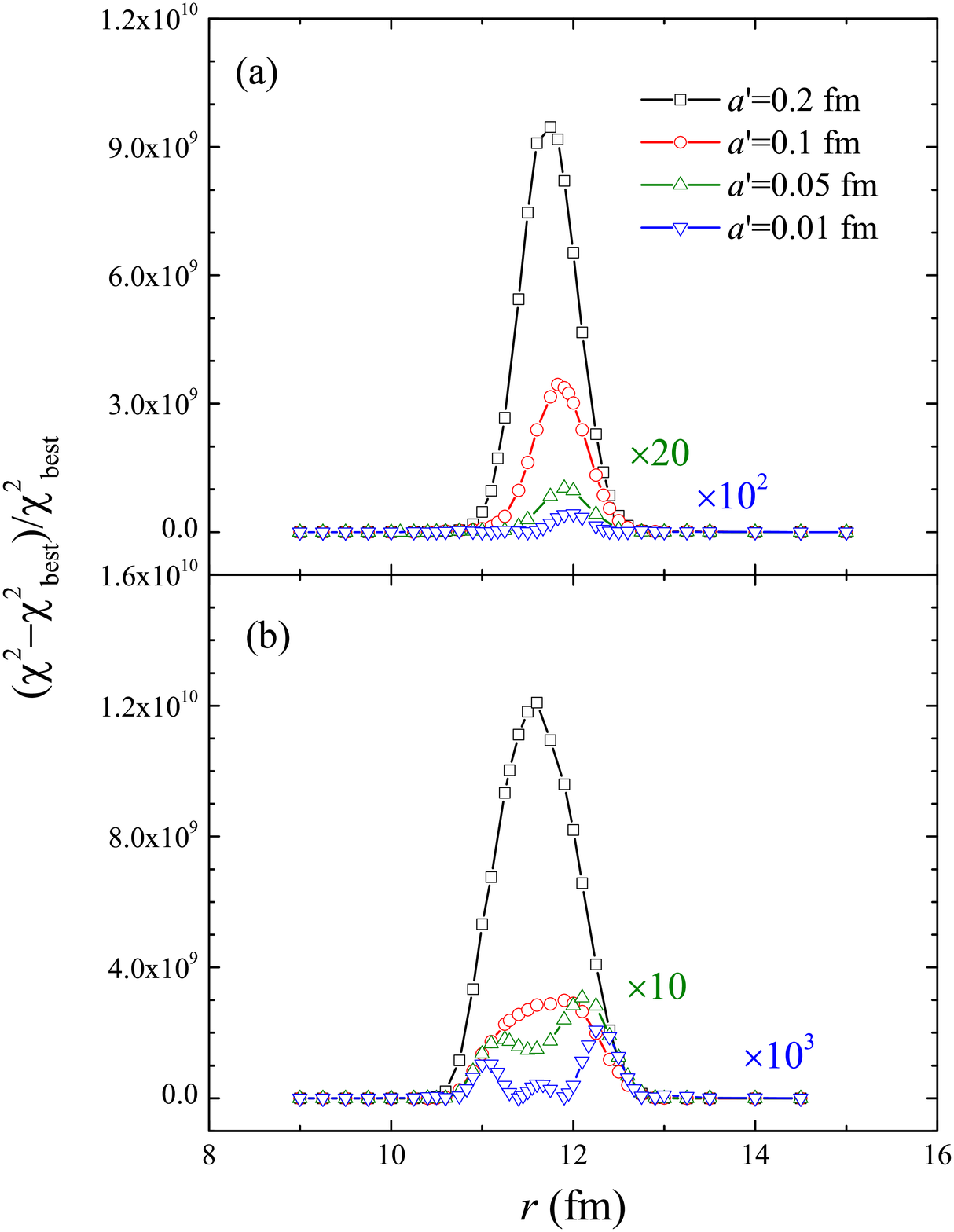}
\figcaption{\label{Fig4}   (Color online) The same as Fig.~\ref{Fig3}, but with different perturbation
width $a'$ values.}
\end{center}

\subsubsection{OMP parameters}

  With the fixed perturbation parameters, further investigations were performed
on the OMP parameters. First, the ideal angular distribution was fitted
with $r_{0i}$ fixed at 1.20, 1.25 and 1.30 fm, respectively,
and results are shown in Fig.~\ref{Fig5}.
Second, the fitting procedure was repeated but with $a_i$ fixed at 0.60, 0.65 and 0.70 fm, respectively,
and results are shown in Fig.~\ref{Fig6}. One can see that relative $\chi^2$ values between the main (major)
peaks and tiny (minor) peaks for the real (imaginary) part vary obviously with the $r_{0i}$ and $a_i$.
However, nearly same sensitive regions were determined by those OMP parameter sets, although the relative
sensitivity differed from each other. Therefore, it can be concluded that the sensitive region determined
by the notch technique is nearly model-independent.

\begin{center}
\includegraphics[width=7cm]{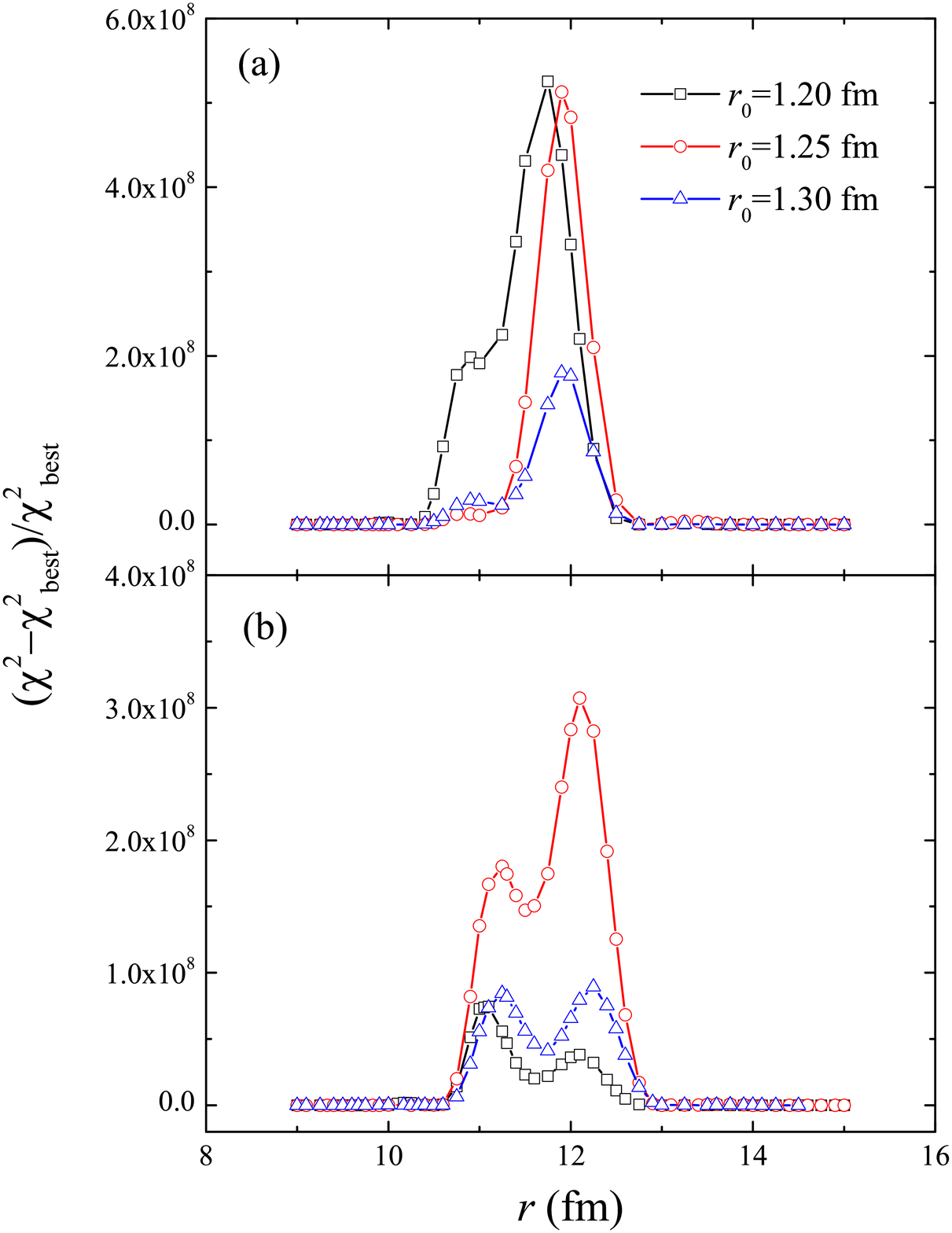}
\figcaption{\label{Fig5}   (Color online) The sensitivity functions for the real part (a) and
imaginary part (b) potentials with different OMP parameters
derived with fixed $r_{0i}$. The curves are used to guide the eyes.}
\end{center}

\begin{center}
\includegraphics[width=7cm]{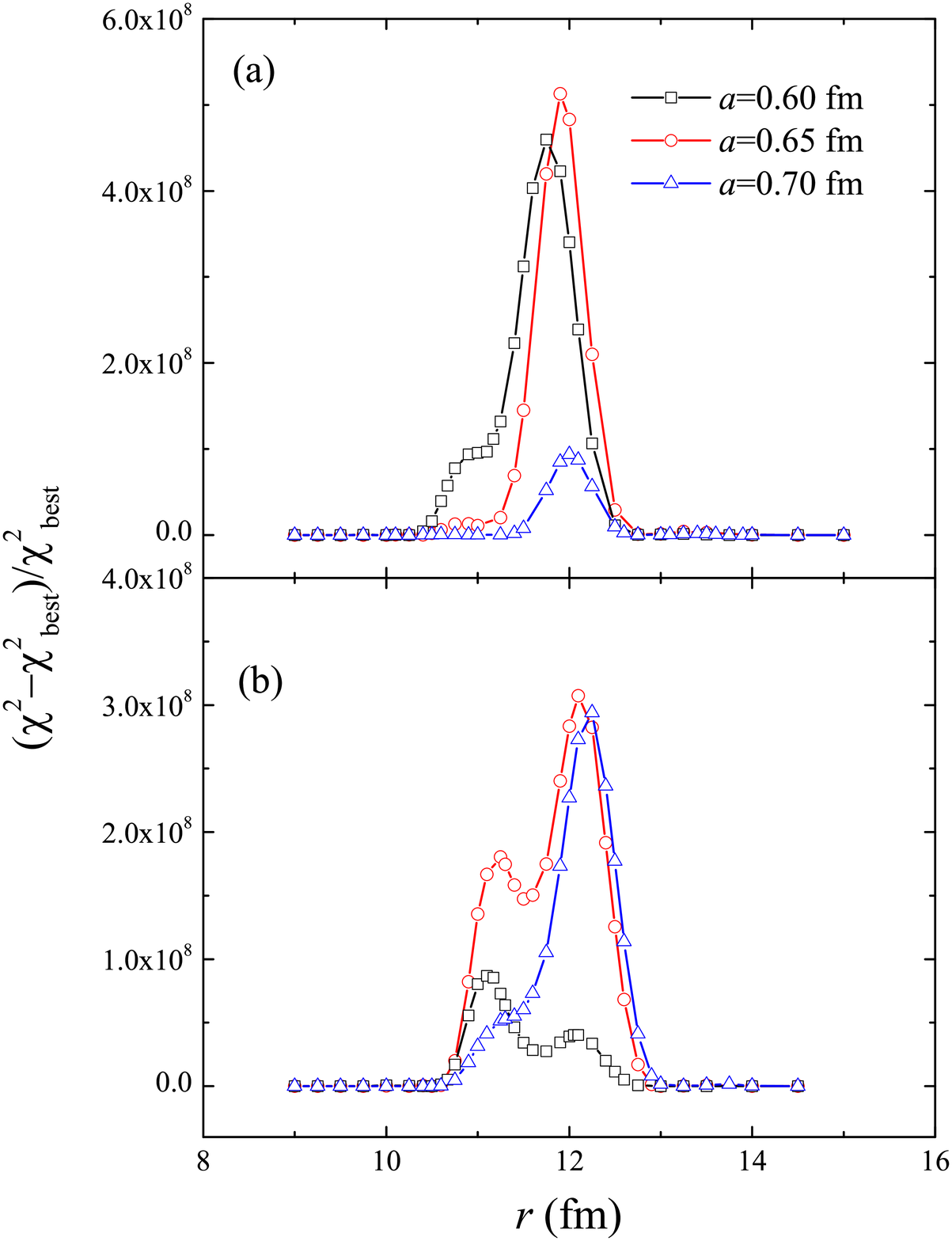}
\figcaption{\label{Fig6}   (Color online) The same as Fig.~\ref{Fig5}, but with the OMP derived
with fixed $a_i$.}
\end{center}

\subsection{\label{sec:level2}dependence on the experimental data }

  After the investigation of the model dependence, a further sensitivity test on the
experimental data was performed, to assess the influence arising
from the quality of the data set, and provide some guidance on the experimental measurements.

\subsubsection{ angle interval}

\begin{center}
\includegraphics[width=7cm]{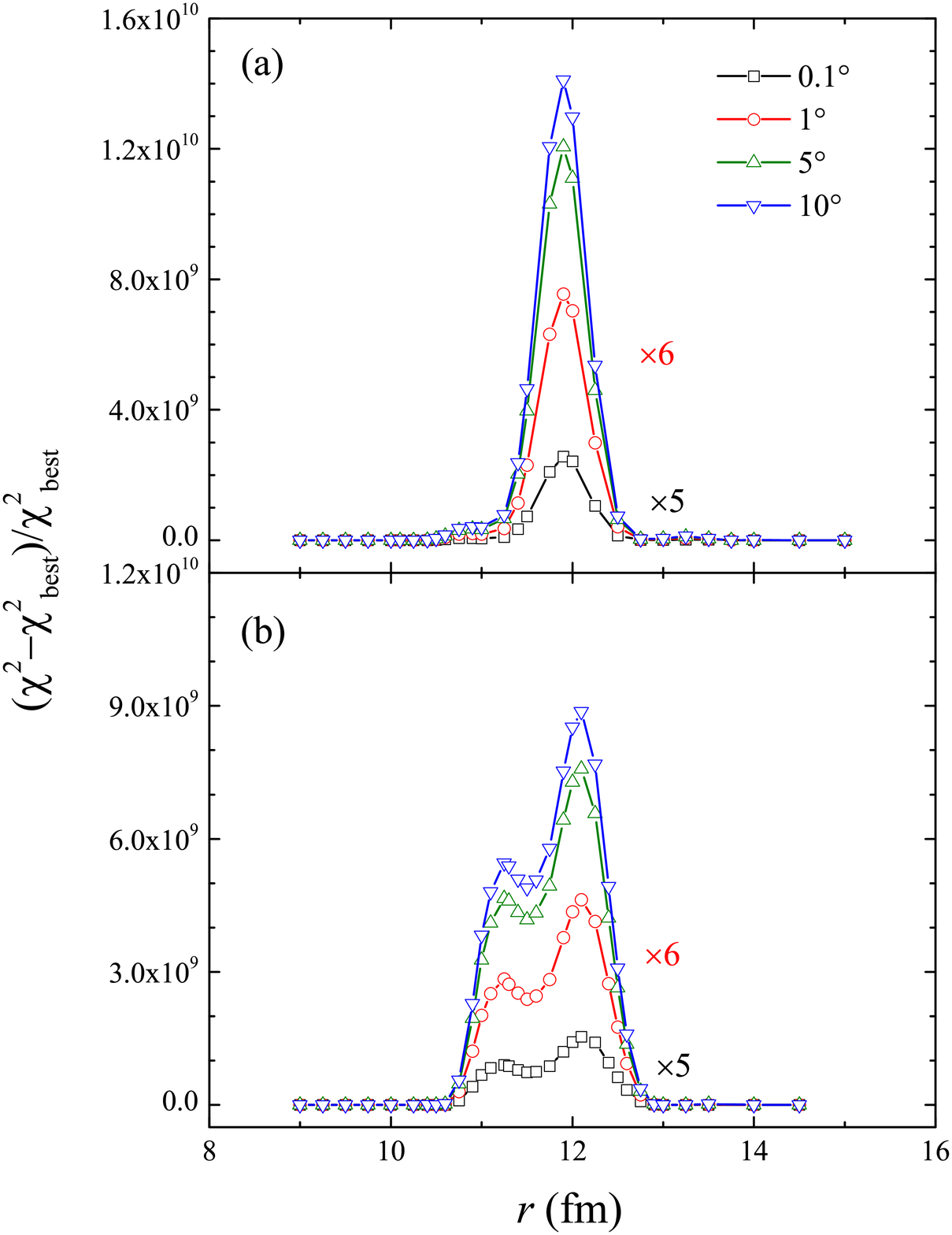}
\figcaption{\label{Fig7}   Color online) The sensitivity functions for the real part (a) and
imaginary part (b) potentials for experimental data set with different $\theta_{\mathrm{int.}}$. The curves
are used to guide the eyes.}
\end{center}

\begin{center}
\includegraphics[width=7cm]{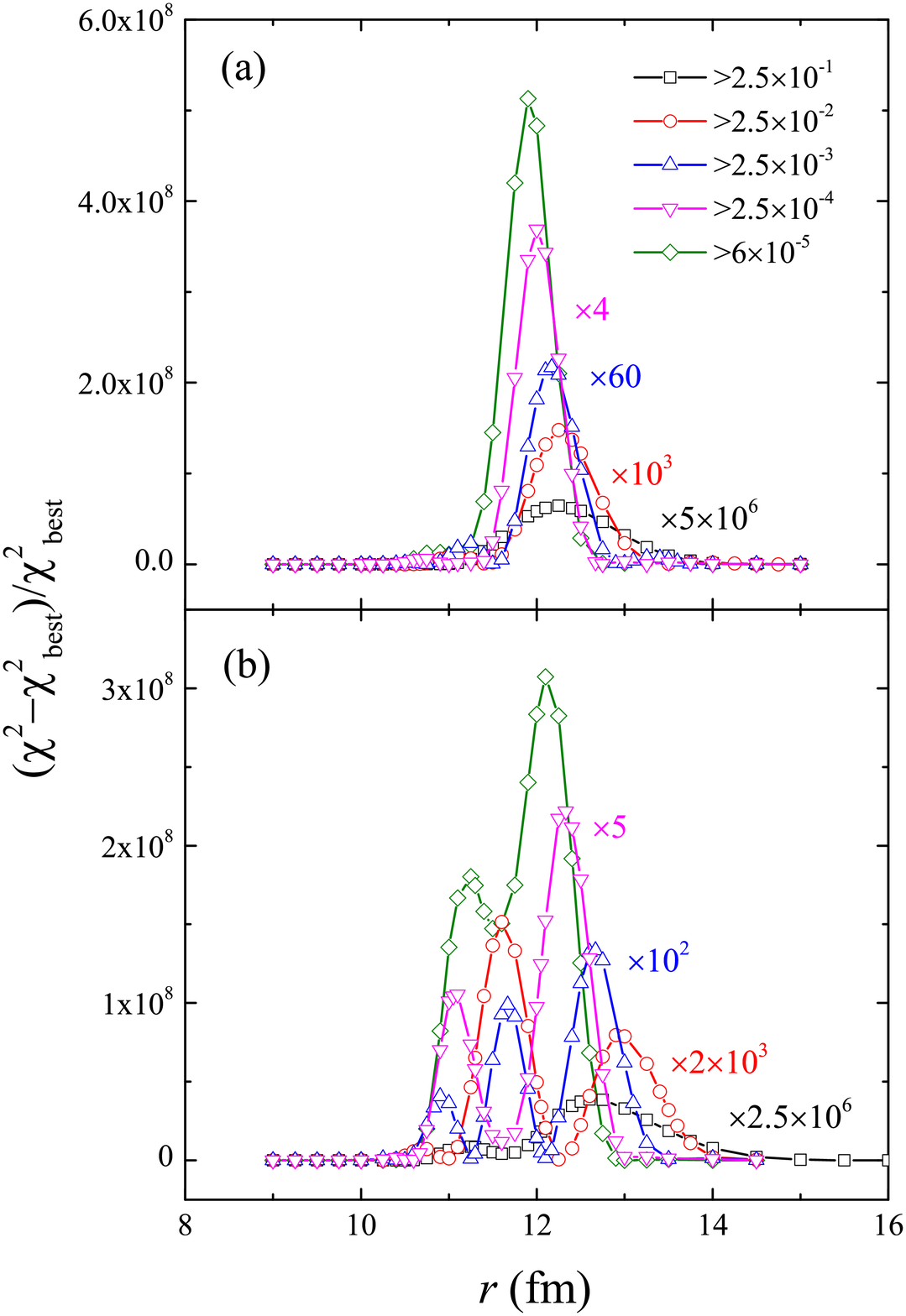}
\figcaption{\label{Fig8}   (Color online) The same with Fig.~\ref{Fig7},
but with different $d\sigma_{\mathrm{el}}/d\sigma_{\mathrm{Ru}}$ extensions.}
\end{center}

\begin{center}
\includegraphics[width=7cm]{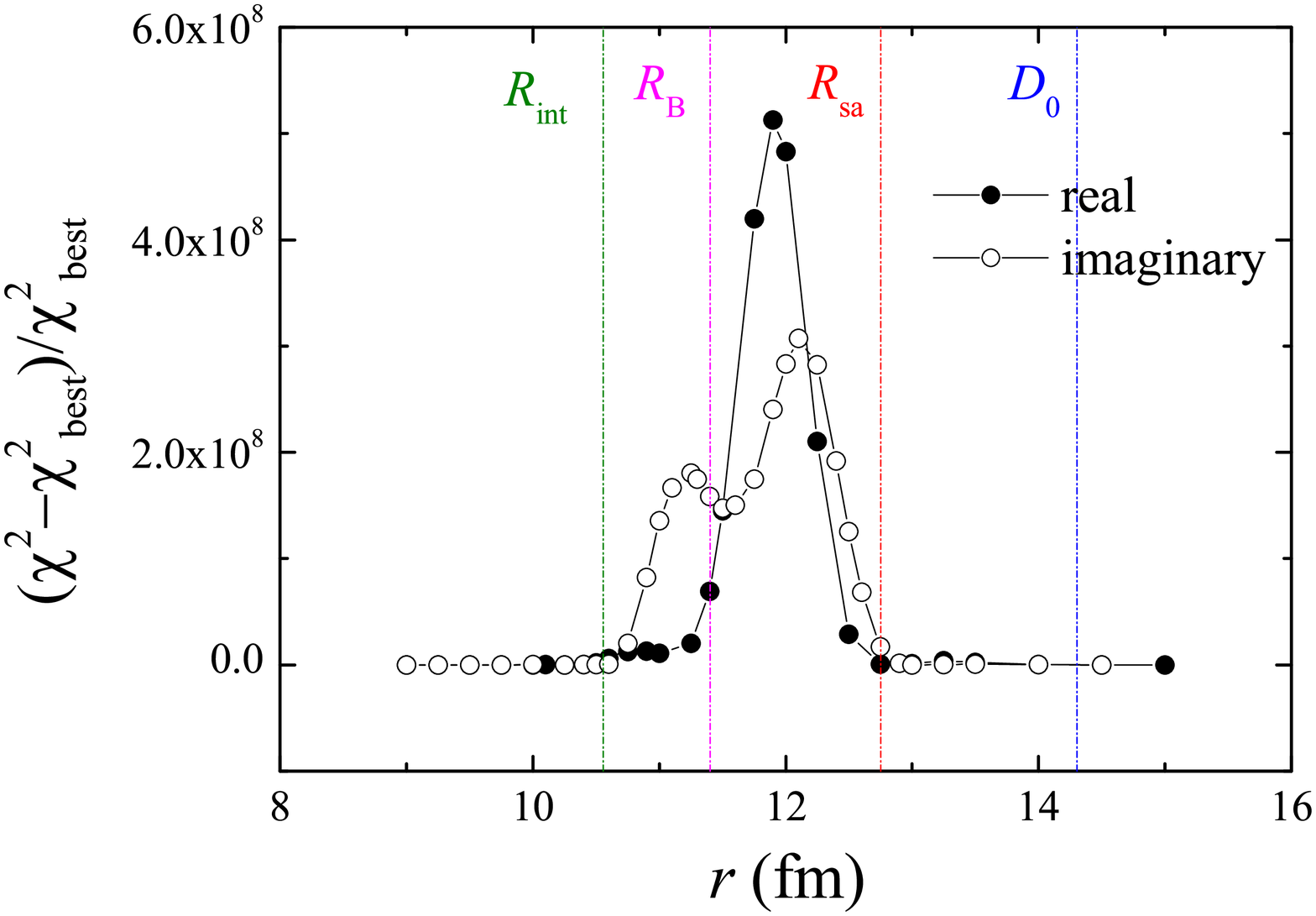}
\figcaption{\label{Fig9}   (Color online) The sensitivity functions of the real (full circle)
and imaginary (hollow circle) parts with the data range down to
$d\sigma_{\mathrm{el}}/d\sigma_{\mathrm{Ru}}=6.0\times10^{-5}$. The interaction radius
$R_{\mathrm{int}}$, Coulomb barrier radius $R_{\mathrm{B}}$,
strong absorption radius $R_{\mathrm{sa}}$, and the distance $D_0$
are showed by vertical lines. See the text for detail.}
\end{center}

\begin{center}
\includegraphics[width=7cm]{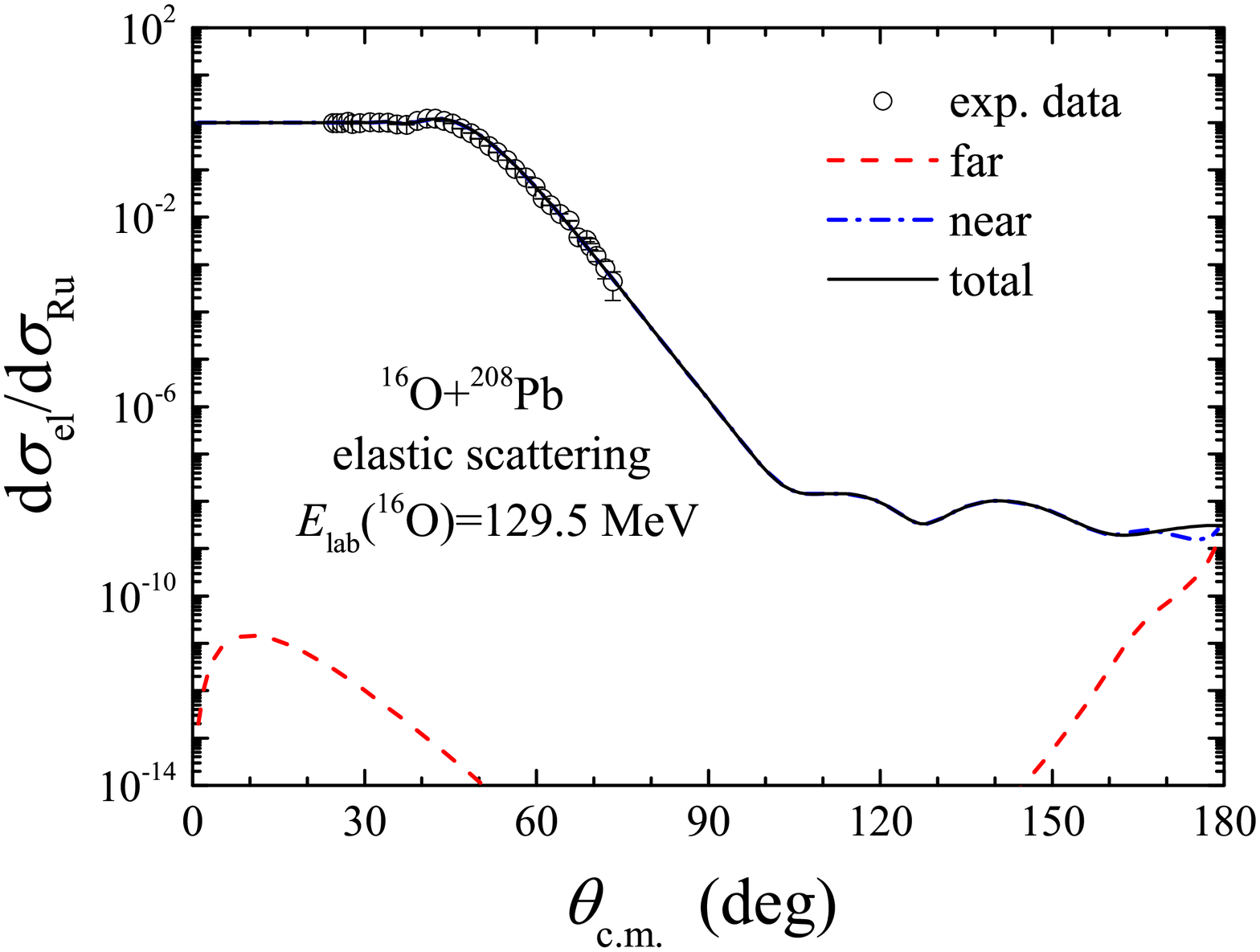}
\figcaption{\label{Fig10}   (Color online) The elastic angular distribution of $^{16}$O+$^{208}$Pb at
$E_{\mathrm{lab}}$($^{16}$O)=129.5 MeV. Open circles represent the experimental data. Best-fit
result is shown by the solid curve. Decomposition of the far- and near-side are also shown by
the dashed and dot-dashed curves, respectively.}
\end{center}

  It can be imagined that a fine angle interval may be useful in determining
a reliable OMP parameter set but it is time consuming,
especially for the RIB experiment. In order to check its influence on the
sensitive region, different angle intervals
$\theta_{\mathrm{int.}}$, i.e. $\theta_{\mathrm{int.}} =0.1^{\circ}, 1^{\circ},
5^{\circ}$ and $10^{\circ}$ were adopted to the ideal data set.
The corresponding sensitivity functions are shown in Fig.~\ref{Fig7}.
It can be seen that bigger $\theta_{\mathrm{int.}}$
introduces a larger $\chi^2$ value. There are no obvious changes in the structures
of the sensitivity functions for both the real and imaginary parts. It demonstrates
that on the premise of large angle-region as well as good statistics, the sensitive
region can be determined accurately even by a few experimental data points. This
conclusion is important to the elastic scattering measurements with
RIBs, whose angular distributions usually have a few points due to the limits of the
intensity and quality of the available RIBs~\cite{Aguilera2001,Aguilera2009}.
However, it should be kept in mind this indication is only available for the heavy target
system, whose angular distribution of elastic scattering is almost structureless.
When refers to the light nuclear system, whose elastic scattering angular distribution
presents strong interference pattern, a fine measurement is necessary to describe the detailed
structure of the angular distribution.

\subsubsection{ angle region}

  In principal, the wider angle region is measured, the more constraint on the OMP
parameters can be achieved. However, it is hard to extend the experimental data to
the large angle where the cross section become very low for the elastic scattering,
especially at high energies. From the physics point of view, data at the back angle
may provide more information on the inner potential. In order to inspect the influences
of angle region measured, the ideal data were divided into five sets:
the first set contains the data down
to $d\sigma_{\mathrm{el}}/d\sigma_{\mathrm{Ru}}=0.25$ at the grazing angle, corresponding to the one-half of the
transmission coefficient; the second set contains the data down to $d\sigma_{\mathrm{el}}/d\sigma_{\mathrm{Ru}}=0.025$,
etc., until the fifth set, which contains all the data points cut off at $\theta_{\mathrm{c.m.}} = 80^{\circ}$, where $d\sigma_{\mathrm{el}}/d\sigma_{\mathrm{Ru}}=6.0\times10^{-5}$. Results of sensitivity test for each data set
are shown in Fig.~\ref{Fig8}. Distinct peaked structures are observed for both the real and imaginary
potential parts. In order to evaluate quantitatively the influences of the data region measured,
the main peak of the real part and major peak of the imaginary part were fitted by a Gaussian
function, respectively. Values of the center position as well as its sigma width are listed
in Table~\ref{Tab1}.

\begin{center}
\tabcaption{ \label{Tab1}  The center(sigma) values of the peaks observed in Fig.~\ref{Fig8}. The real
and imaginary correspond to the main peak of the real part and the major peak of the imaginary part,
respectively. The value of the center is in the unit of fm.}
\footnotesize
\begin{tabular*}{80mm}{c@{\extracolsep{\fill}}ccc}
\toprule data range & real   & imaginary \\
\hline
$>2.5\times10^{-1}$&   12.31(0.56)	&12.76(0.57)\\
$>2.5\times10^{-2}$&  12.29(0.34)	&13.02(0.34)\\
$>2.5\times10^{-3}$&	12.17(0.25)	&12.68(0.25)\\
$>2.5\times10^{-4}$&	12.01(0.23)	&12.33(0.25)\\
$>6.0\times10^{-5}$&	11.91(0.25)	&12.10(0.31)\\
\bottomrule
\end{tabular*}
\end{center}

  For the first data set only containing the data with
$d\sigma_{\mathrm{el}}/d\sigma_{\mathrm{Ru}}>0.25$, the nuclear force just begins to
take effect, thus it is difficult to obtain the accurate information of the nuclear potential,
so a very broad peak presents.
As we extended the data to larger angles, the effects of the nuclear force begin to increase,
and more detailed information can be extracted. Meanwhile both the main
peak of the real part and the major peak of the imaginary part systematically
move into the inner with angle going backward. However, considering the peripheral
nature of elastic scattering, the inner potential can not be probed when the distance
is short than a certain value.

  Moreover, one can find that the lower cross section of the elastic scattering
is measured, the larger relative $\chi^2$ value will be got. The relative $\chi^2$
value of the fifth set is $10^{7}$ times larger than that of the first set. As mentioned above,
the value of relative $\chi^2$ represents the sensitivity degree of the OMP parameters
on the elastic scattering data. Such a large relative $\chi^2$ value demonstrates that
a well adequate constraint can be achieved for the OMP parameters within the sensitive
region when the measurement reaches to a very low cross section.

\subsection{\label{sec:level2}discussion }

  The ideal test provides a solid foundation for the application of the notch technique.
With the appropriate parameters of the notch and OMP, the physical meanings of sensitivity
peaks can be understood. In order to demonstrate clearly, several available radii and distance, e.g. the
radius of the interaction potential $R_{\mathrm{int}}$, Coulomb barrier radius $R_{\mathrm{B}}$,
strong absorption radius $R_{\mathrm{sa}}$, as well as the distance $D_0$ where the nuclear
force begins to take effect, are labeled in Fig.~\ref{Fig9} by vertical lines.
The $R_{\mathrm{sa}}$ is the radius where the observed cross section has fallen to
one-fourth of the Rutherford value; and $D_0$ corresponds to the distance of
$d\sigma_{\mathrm{el}}/d\sigma_{\mathrm{Ru}}=0.98$. One can find that even for data down to
$d\sigma_{\mathrm{el}}/d\sigma_{\mathrm{Ru}}=6.0\times10^{-5}$, the main sensitivity regions
located around 12.0 fm, far larger than the $R_{\mathrm{int}}$, 10.56 fm,
demonstrating that the OMPs determined by the elastic scattering are only sensitive to surface regions.
And as mentioned in Ref.~\cite{Moffa1976}, because of the strong absorption, it seems
unlikely that much light can be shed on the behavior of the real potential in the deep
interior region with measurements of heavy-target system elastic scattering.

  For the imaginary part, two distinct peaks are observed. The major one lies around the
$R_{\mathrm{sa}}$, corresponding to the surface absorption process; the minor peak locates
around the $R_{\mathrm{B}}$, which should be responsible for the volume absorption, i.e. the
capture reaction process. While for the real part, the main peak locating near the $R_{\mathrm{sa}}$,
followed by a tiny inner peak, which lies inside of the $R_{\mathrm{B}}$. Both of the two real-part peaks
locate inside of the corresponding imaginary ones. The main peak of the real part is
arising from the direct scattering process. The origin of the tiny peak was thought
to be associated with the far-side interference effect~\cite{Cramer1980,Roubos2006}.
In order to check the reliability of this explanation, the decomposition of the
far- and near-side scattering was performed with
the method developed in Ref.~\cite{Anni2002}, and the result is shown in Fig.~\ref{Fig10}.
One can find that the far-side scattering is almost negligible for the whole angle range,
indicating that the tiny peak should not be originated from the interference between the far- and
near-side components. Considering that the location of the tiny peak is inside of $R_{\mathrm{B}}$,
we believe this peak should be the result of the resonance scattering, where the compound nucleus has
been formed.

\section{\label{sec:level1}Summary and conclusions}
  The sensitivities of the notch technique on the parameters of the perturbation and OMP,
as well as the experimental data were investigated in the present work. Through the ideal
test we can draw the conclusions as below:
  1) a sufficient large reduced fraction $d$ can help to obtain accurate information of the
sensitive region, and $d$=1.0 is the adopted value;
  2) the width of the perturbation $a'$ should be several times of the integration step $\mathrm{d}r$,
an inappropriate large or small value of $a'$ will lead to a spurious result;
  3) the notch technique is almost independent on the OMP parameters;
  4) for the heavy-target nuclear system, on the premise of large angle-region as well as good statistics measurement,
there is no need of a great deal of experimental data points
to ensure the reliability of the sensitive region extracted. This may aid in
optimizing the setup of elastic scattering measurement, especially for the experiments with RIBs;
  5) the relative inner information of the nuclear potential can be derived when measurement extended to a
lower elastic scattering cross section. However, the deep interior region of the nuclear potential
is still invisible through the elastic scattering measurement due to the effect of strong absorption.

  With these detailed investigations of the notch technique, we can further apply this method
to the researches on the radial sensitivities of both tightly- and weakly-bound nuclear systems,
which are essential issues in the studies of the OMP.

\end{multicols}

\vspace{10mm}

\begin{multicols}{2}

\end{multicols}

\clearpage

\end{document}